# Simulating Behaviours to face up an Emergency Evacuation

Pablo Cristian Tissera, Alicia Castro
LIDIC-Departamento de Informática

Universidad Nacional de San Luis
Argentina
Email:{ptissera,adcastro}@unsl.edu.ar

A. Marcela Printista
LIDIC-Departamento de Informática

Universidad Nacional de San Luis
Conicet CCT-San Luis, Argentina
Email: mprinti@unsl.edu.ar

Emilio Luque
Departamento de Arquitectura y
Sistemas Operativos
Universidad Autónoma de Barcelona-
España
Email:emilio.luque@uab.es

*Abstract* Computer based models describing pedestrian behavior in an emergency evacuation play a vital role in the development of active strategies that minimize the evacuation time when a closed area must be evacuated. The reference model has a hybrid structure where the dynamics of fire and smoke propagation are modeled by means of Cellular Automata and for simulating people's behavior we are using Intelligent Agents. The model consists of two sub-models, called environmental and pedestrian ones. As part of the pedestrian model, this paper concentrates in a methodology that is able to model some of the frequently observed human's behaviors in evacuation exercises. Each agent will perceive what is happening around, select the options that exist in that context and then it makes a decision that will reflect its ability to cope with an emergency evacuation, called in this work, behavior. We also developed simple exercises where the model is applied to the simulation of an evacuation due to a potential hazard, such as fire, smoke or some kind of collapse.

*Keywords— Evacuation Simulation; Behaviors; Cellular Automata; Intelligent Agents*

## I. Introduction

In the last years, several modeling approaches have been proposed to deal with the emergency evacuations because the prediction of the people's behavior is of great public interest. Microscopic approaches allow to investigate how the system state evolves during the model runs. Within this category, there is a diversity of models, including perhaps the best known are those based on the Social Forces Model (SFM) [1], Cellular Automata (CA) [2], [3], [4], [5] and Multi-Agent System (MAS) [6], [7]. Models based on the concept of social forces and cellular automata, can represent individuals as basic units of the system but they have the limitation of assuming that the individuals depicted are homogeneous, i.e. the behavior of all of them will be governed by the same set of rules [8].

CA allow to generate "local" and "uniform" behaviors that resemble the dynamics observed in real processes of fire and smoke propagation. CA can be viewed as a simple model of a spatially extended decentralized system made up of a number of individual components (cells). The communication between their cells is limited to local interaction. Each individual cell is in a specific state which changes over time depending on the states of its local neighbors.

However, these local features were not suitable for representing certain aspects of people's behavior that requires a more specific and differentiated perspective. We developed a hybrid model where the dynamics of fire and smoke propagation are modeled by means of CA and for simulating people's behavior we are using intelligent agent (IA) concept.

In order to cover a variety of scenarios suitable for checking safety issues, in this model we will focus on pedestrian behavior. In the model, each individual or pedestrian is an agent that will have its own thread of control and be able to run appropriate actions according to their own state, the perceived environment and the messages provided by the system (external stimuli) independently.

The proposed strategy is similar to the philosophical theories of Dennett and Bratman [9], [10], [11] where each agent will create a high-level model with actions to be taken to achieve a goal and plan how to implement them. The agent perceives what is happening around, selects the options that exist in that context and then the agent makes a decision that will reflect its ability to cope with an emergency evacuation, called in this work, behavior. Thus, in the model creation are involved three sub-processes, where different configurations of them carry out different scenarios for the agent.

Our model considers that the making-decision process that an agent makes to evacuate will be driven by an engine of behavior which concentrates different mindsets or reactions to face up to an emergency situation.

Through interaction and coordinated evolution of CA and IA models, it is possible to obtain a model capable of simulating indoor environments with a finite number of exits that must be evacuated by individuals with different behaviors due to an emergency evacuation. The proposed simulation system allows specifying different scenes with a large number of people and environmental features, making easier the study of the complex behaviors that arise when the people interact. Our proposal could be used by architects, government agencies, foundations, etc. in order to know the security threats of a possible disaster, help with appropriate actions of prevention for in a quick and efficient way to evacuate a building through the design of active policies that minimize the evacuation time when circumstances require it.

In section II we present the Hybrid Model for the Evacuation Simulation. Sub-section II-A resumes the sub-model called environmental (EsM). Sub-section II-B explains the pedestrian sub-model (PsM) and addresses the agent architecture adopted in this work. The section III explains the implementation of two behaviors commonly observed in emergency evacuation. In section IV we describe our work with different instances of the problem at hand and report the performance analysis of each case. Finally, the section V presents the conclusions and the section VI we put forward possible future works.

## II. HYBRID MODEL

The use of cellular automata (CA) is attractive because it is a simple and extremely efficient model for computer simulation. Unfortunately, these models impose certain limitations that make it difficult the emergence of different social behaviors, from the interaction of a heterogeneous group of individuals with their perceived environment.

The considered simulation model is able to take advantage of both cellular automata and intelligent agents [12]. It consists of two sub-models, called environmental (EsM) and pedestrian (PsM). This model along with the computational methodology allows us to build an artificial environment populated with autonomous agents, which are capable of interacting with each other. Fig. 1 shows the hybrid model.

The EsM, based on CA, describes the spatial configuration of the environment (geometry of space, exit doors, internal barriers, etc.) and models the processes of diffusion of smoke and fire, while the PsM uses the concept of intelligent agents to describe the cognitive processes of individual agents and interactions among multiple agents in a specific environment. Through interaction and coordinated evolution of these two sub-models it is possible to obtain a model capable of simulating indoor environments with a finite number of outputs that must be evacuated by a group of people due to the threat of fire and the effect of the smoke. The evacuation exercises adopt the CA model for advancing the simulation time.

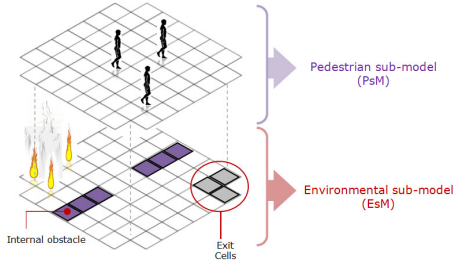

Fig. 1. Hybrid Model consisting of environmental and pedestrian sub-models.

### A. CA Model for Environmental Simulation

The CA are discrete dynamic systems with capacity to develop complex behaviors from a simple set of rules. Basically, these rules allow specifying the new state of a cell based on the state of the neighboring cells. In this way, it is possible to model complex dynamic systems from the specification of the local dynamics of its components. The Cellular Space is a finite bi-dimensional array (grid) with closed boundaries. Each cell of the cellular space represents $40 \times 40$ cm$^2$. Each cell can be in one of the states of the set Q= {W, E, P,O, S, SF, PS}, where:

| W: External wall cell | E: Empty cell | S: Cell with smoke |
|---|---|---|
| P: Cell with a person | PS: Cell with a person and smoke | |
| O: Internal obstacle cell | SF: Cell with smoke and fire | |

The neighborhood considered in the model is *Moore's Neighborhood*, that includes the eight cells surrounding the central cell. With this choice we aim to provide to each individual in the system with all possible movement directions. Before the simulation starts, a variety of information related to the outer walls, inner obstacles, individuals, combustible locations, cell with fire and arrangement of the exits must be specified. In our tools this task can be done by means of its graphical interface.

*Model Evolution Rules*: a cell in state W or O (outer wall or obstacle) will not change its state throughout the simulation (Rules about the building). For fire and smoke spreading we use a probability-driven model [13]. A cell with smoke (*S, SF* or *PS*) at time *t*, will also have smoke in time *t+1*. If at time *t* the central cell does not have smoke, but some of its adjacent cells have smoke, the central cell will also have smoke at time *t+1* with a proportional a probability to the number of adjacent cells with smoke. The rules about *fire propagation* are analogous to the rules about *smoke propagation*. If a cell has smoke or fire, it will change its status from a *min-value* that indicates the recent emergence of the element (smoke or fire) in the cell, to a *max-value* indicating that the cell has been fully saturated by the spread element (smoke or fire).

The simulation takes an updating time of 0.3 seconds by time-step. This value is the estimated time required by a pedestrian for walking 0.4 m (size of a cell side).

For reasons of space we only include a summary of the CA model, for details to refer to previous work.

### B. Pedestrian Model

Intelligent agents (IA) have been used successfully in a wide range of applications. In artificial intelligence, an intelligent agent is an autonomous entity which observes and acts upon an environment and directs its activity towards achieving goals. The PsM is the part of the hybrid model focuses on representing the human behaviors. Evacuation processes can be considered a special type of crowd dynamics, with the presence of hazards. Each agent has a set of static characteristics (such as sex, age and level of stress tolerance) and dynamic (such as current stress level and speed). In addition to that, the crowd behaviors are complex phenomena, it is necessary to focus on various factors that determine the human behavior and that may contribute to emergency behaviors.

In this paper, the agent behavior will be determined by their current perceptions and attitudes of reaction. This type of system provides solutions in dynamic and uncertain environments, where the agent has only a partial view of the problem. The procedure can be followed in the Algorithm II.1.

At first, the agent must observe the environment and gather the state of outside world with its inner world (step 1 of the algorithm II.1). With this information the agent updates its knowledge and starts to think about their possibilities of escape (step 2). Consider for example two types of individuals: a person familiarized with the place (*kp*) and another that completely ignores the place to evacuate (*ukp*). The agent *kp* has a rich set of information to built their alternatives, whilst the world of *ukp* just contains what it is watching right now and what it was able to learn until evacuation began. After step 2, they will have different knowledge.

Next, the agent should contrast their alternatives before making a decision. Both Dennett's theory about intentional systems, as well as Bratman's theory of practical reasoning about humans, agrees that in any act of a rational agent at least involves a deliberative process and a process of practical reasoning leading to an action.

This deliberative process will take into account both the world and the agent options. These latter will be biased by its personality type and behavior. Will it choose the nearest exit? will it choose the exit that it infers faster? The agent gets lost, will it follow the other agents? The different behaviors that an agent can adopt represent their options. In our implementation, each agent has an engine of behaviors allowing it to have different goals based on their different escape criterion. The engine of behavior manages deliberative process (step 3). As can be seen in Fig. 2, it is a non-deterministic finite automaton, where each node represents the implementation of a behaviors while the transitions represent the event for which the individual can change its behavior. Now, the deliberative process concerns with combining new facts with existing knowledge for solving different situations. At the end of the deliberation process the agent will get its objective, that consists of the selection of a specific way to evacuate ("the agent decided to behave this way") and it will commit with it. For example, the agent *ukp* could have decided to evacuate by the nearest exit, then it must select the closer exit regardless of the number of people heading to the same exit.

Finally, the agent somehow must get the selected exit (its plan) (step 4). The plan will consist in following a set of paths that lead to the selected exit. In our implementation, the set of paths is determined by Dijkstra's algorithm from the cell occupied by the agent to the selected exit.

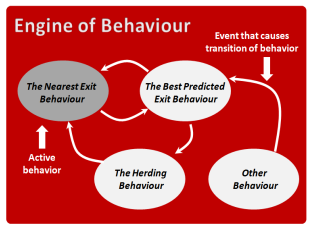

Fig. 2. The deliberative process implemented by an engine of behaviors.

The decisions that the agent is taking are continually adjusted by a process that takes into account both characteristics of the individuals and the dynamics of their perceptions. So, the agent does the following:

1) If its plan is reasonable (step 9), i.e. the agent can achieve the objective with the current plan, then it will execute the next step of the plan (step 6), otherwise the agent should select a new plan (step 10). For example, on the way to an exit, an obstacle appears and then the agent must find another route to the objective exit.

2) If the agent decides that its alternative is not neither feasible nor desirable (step 5) then it will return to the outer loop and deliberate about the convenience of select another objective (another exit) or even to change of strategy, resulting in this last case a change of state in its engine of behavior (step 3). For example, an exit can be blocked completely by the spread of the fire along the evacuation evolution; therefore there will be no possible path that leads to it.

**Algorithm II.1**: control loop of the agent()

**while** (NOT(success evacuation))

**do**
- Get next world state (1)
- Update knowledge (2)
- Deliberates and Commits with an objective (3)
- Get plan to achieve the objective (4)

**while** (NOT (empty plan or success evacuation or feasible objective)) (5)

**do**
- Execute plan (6)
- Update knowledge (8)

**If** NOT (reasonable plan) (9)
**Then** - Get new plan (10)

**End**.

### III. BEHAVIORS

This section explains the process for two types of individual behaviors implemented till now. In the first behaviors, called the Nearest Exit, the agent will try to get out the closest exit to its current position. In this behavior the decision process will take into account the position of the agent, the direction toward the nearest exit, the state of its environment in relation to the progress of fire and smoke, but it ignores information from other alternative solutions, the behaviors of other agents and it will not take unexpected or altruistic decisions. The decision-making process selects the exit with the shortest $PredDist_j$. This measure represents the estimated distance from current position to the exit $j$.

In the second behaviors, the agent will analyze different exits and choose one that it predicts the fastest exit to evacuate. In this behaviors, called the *Best Predicted Exit*, the decision process will consider the position of the agent, the state of its environment in relation to the progress of fire and smoke, the distance to alternative exits, the density of the crowd trying to evacuate for each exit (only if the agent can see the exit) and the stress level in relation to its tolerance to it. Under this behavior, the agent does not follow the actions of others, i.e. it remains in the category of individual behavior but it only checks the orientation of the other agents to choose its way toward the best exit. As the evacuation progresses, the agent is predicting the cost (in time) to evacuate by each of the exits that are available in the environment. The inferred lower cost will indicate the best exit. For that, the decision-making process evaluates the following cost function:

$Cost_j = MinPredTime.To_j * PredDist_j * I_j;$
   if $MinPredTime.To_j >= EvacPredTime_j$.

In other case:
$Cost_j = EvacPredTime_j * MinPredTime.To_j * PredDist_j * I_j;$

where:

- $I_j$ represents the estimated number of agents who intend to evacuate through the exit j ;

- $EvacPredTime_j$ represents the evacuation estimated time of exit *j*. This factor takes into account the intentions of other agents to escape by the door *j*;

- $MinPredTime.To_j$ represents the minimum estimated time needed by the agent to reach the exit j considering a free path (unobstructed); and

- $PredDist_j$ represents the estimated distance from current position to the exit *j*. This calculus is made using the Dijkstra algorithm.

Note that for the case where $MinPredTime.To_j >= EvacPredTime_j$ it does not take into account the value of $EvacPredTime_j$, because if the spent time by the agent to reach the door "*j*" is greater than the time taken to dislodge the door, then when the agent arrives at the door, it would find it empty.

The resulting procedure instructs the agent for which exit to go (*objective*). In addition, the procedure involves a dynamic factor used to adjust the number of times the agent executes the action to evaluate the exits. This factor limits the effect of indecision of the agents that occurs when simultaneously multiple agents make the same prediction of fastest exit to evacuate. As a result, as time goes on, agents can questioning the decision taken by themselves and try to execute the deliberation process again (the agent comes and goes). This factor depends on the environment size and it is dynamically adjusted according the time elapsed since the beginning of the evacuation.

*Path to the Selected Exit*

Each agent knows the objective exit. Based on the perception of its environment, the agent must select at each time step, a position on the path to its exit.

If in the neighborhood there are free cells that come near the exit, the agent will choose the one that most benefits grant (leave it closer). If there is more than one, the choice of the position is random. If in the neighborhood there is no free cell, following the approach described above, the agent will select a cell that is currently occupied by another agent. This can cause multiple agents try to occupy the same physical location belonging to an exit way. To solve the problem, after the agents expressed their desire to move to the nominated cell, they should delay their movement until a conflict resolution process is executed. To solve the collisions we changed the approach commonly used in other works. Instead of being the agents who decide, the selected cell is carrying out the agents competition process. The hosting decision is concentrated in each nominated cell.

The conflict resolution process must solve two types of situations:

- The conflict occurred when multiple agents chose the same *free cell*. In this case the process gives priority to the selection of agents with greater speed and fewer points of damage (parameter specified in the environment). If the conflict persists, the selection will be random.

- The conflict occurred when multiple agents requested a *cell occupied by another agent*. The process must check if the cell will be free in the next time step. If it will be free, the procedure of the same free cell is executed. Otherwise, the agent will not move of its current cell.

Finally, it is important to point out that after a prudential time, if an agent remains without advancing, it will try to change its neighborhood. In spite of failing to advance to the desired direction, it is reasonable to want to keep moving until finding a way towards the exit.

## IV. TEST-CASE SCENARIOS AND RESULTS

The experiments were carried out with EVAC Simulator [2], an integrated simulation system. EVAC is a system developed in Java that allows the design and simulation of spatial environments in an explicit way. EVAC simulator offers a friendly graphical interface which can be easily used by non expert users. We performed a series of experiments in order to test the behaviors under study focusing on showing the results of the interaction of the two sub-models. For all the cases, the environment is of 20 x 30 meters with 625 individuals and two Exits of 1.6 meters each one located in opposite ends of the environment (called $E_1$ and $E_2$). We will call $\beta_{NE}$ to those individuals whose configuration of behavior is Nearest Exit (*NE*) and $\beta_{BPE}$ to those whose configuration of behavior is Best Predicted Exit (*BPE*).

With the purpose of obtaining acceptable statistical data, the results shown in Table I, II and III correspond to the average of 50 independent replications of each experiment. The corresponding confidence intervals were obtained, for the total evacuation time (TET) (seconds), mean evacuation time (METxI) (seconds) and mean travelled distance (MDxI) (meters) per individual to the exit. For the experiments called A and B, all of the individuals in the environment have $\beta_{NE}$ behavior. The cases of study C and D are analogous to the previous ones but they have $\beta_{BPE}$ behavior. It is important to note that in the cases A and C, the environment is not affected by the spread of fire and heat, but in cases B and D individuals must adapt to a dynamic environment due to the spread of fire and heat.

As can be seen in Table 1, in the case A about 60% of individuals decided to use $E_2$. In the case B, a fire started in an area near $E_2$. From the perception of this danger zone, agents were going to that exit began a new deliberation process and some of them could change its objective.

Remember that in cases A and B, people choose the exit that they consider closest to them (for distance). The pedestrian model considers zones of fire, smoke or heat as obstacles (dynamic characteristic of the environment), so paths through that zone will be considered invalid. When fire appears, agents must re-plan their strategy to surround the fire or escape on the other hand. Of the original 365 individuals who had the intention to walk out the $E_2$, 45% of them considered that the $E_1$ was the nearest exit and therefore changed their objective, while the rest of the agents surrounded the fire (by another plan) but maintained their intention to go out the $E_2$. Finally, in this experiment, about 68% of individuals were evacuated by the $E_1$.

This situation also causes an increase in the evacuation times and the mean travelled distance per individual to the exit as shown in Table 1 for cases A and B.

In a similar way we carried out a new series of experiments (cases C and D), but this time the individuals can choose any exit to evacuate according to its perceptions of the environment and a making-decision process to select the exit that the agent predicts faster. In the case C, the emergent behavior of individuals, results in equilibrium in the selection of exits. Approximately 50% of individuals choose to evacuate by $E_2$ while the remaining 50% choose evacuate by $E_1$. In case D is possible to see how the amount of individuals that decided to evacuate for the $E_2$ is decreased due to the threat of fire and the heat in approximately 10% (around 61 individuals). The differences that can be viewed between the evacuation times and the mean travelled distance per individual of the environments A vs. C and B vs. D, lies in that in the environments C and D some individuals, after a process of deliberation, decided to change more than once their objective (Exit through evacuate) along the evolution of the model, i.e. the individuals re-evaluate its decision and obtain a new objective, which may be interpreted as a kind of "repentance". Fig. 3a (right) shows the behavior of individuals to the case A. In Fig. 3b (left), we can see how a lot of individuals who had decided to evacuate by $E_2$ now goes to $E_1$ due to the spread of fire and heat in the environment. Both figures were obtained through the display module of our tool when the evacuation time is equal to 30 seconds.

The second set of experiments (cases E, F and G) has as aim show the existing interaction between both types of individuals. In this set of experiments approximately the half of the individuals has $\beta_{NE}$ behavior, while the remaining individuals are $\beta_{BPE}$. The difference between these three case studies lies in the number of cells that can be achieved by the spread of fire and heat. In the cases F and G the environment changes in a dynamic way due to the spread of fire and heat, but the quantity of affected cells is minor for the case of study F that for the case G. On the other hand, in the case study E the environment is not affected by the spread of fire and heat.

TABLE I: 625 INDIVIDUALS DISTRIBUTED EVENLY IN THE BUILDING. 2 EXISTS. #$E_i$ INDICATES THE NUMBER OF PEDESTRIAN EXISTING DOOR $E_i$

| Case | TET (sec) | METxI (m) | MDxI (m) | #$E_1$ | #$E_2$ |
|---|---|---|---|---|---|
| A | 98.91- 99.56 | 13.40-13.41 | 16.07-16.08 | 262.00 | 363.00 |
| B | 108.28-122.51 | 20.05-20.12 | 24.05-24.15 | 425.94 | 199.06 |
| C | 101.63-106.40 | 14.43-14.44 | 17.32-17.33 | 301.86 | 323.14 |
| D | 118.01-122.30 | 28.17-28.26 | 33.79-33.91 | 363.10 | 261.90 |

The results obtained for this series of experiments show us that there is a decrease of approximately 10% of individuals who select the $E_2$ as objective due to the appearance of fire and heat in the neighborhood of it (Table II cases E and F and exit # $E_2$). Also it is possible to visualize that as there is increase the amount of individuals $\beta_{NE}$ to evacuate by $E_1$, the number of individuals $\beta_{BPE}$ that decide to evacuate for the same exit decreases (Table II). This behavior more risky observed in individuals BPE is due to its perceptions they considered as a factor to define its objective the congestion of the exits. Finally as you would expect both the evacuation times as the mean travelled distance per individual (Table III) are increased due to the need of the individuals to surround the areas with fire and heat.

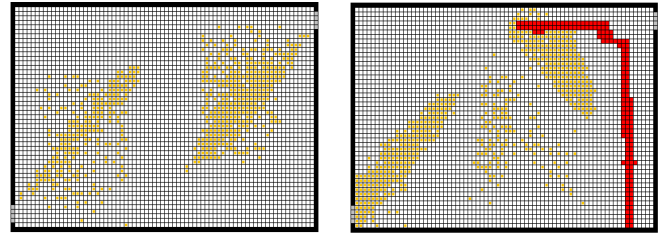

Fig. 3a. (right) Case A. Without fire - 100% $T_{NE}$. $E_1$: down left, $E_2$: up right.
Fig. 3b. (left) Case B. With heat and fire - 100% $T_{NE}$

TABLE II: 625 INDIVIDUALS DISTRIBUTED EVENLY IN THE BUILDING. 2 EXISTS. #$E_i$ INDICATES THE NUMBER OF PEDESTRIAN EXISTING DOOR $E_i$. #$E_i$ $\beta_{NE}$ AND #$E_i$ $\beta_{BPE}$ INDICATE THE NUMBER OF PEDESTRIAN $\beta_{NE}$ AND $\beta_{BPE}$ EXISTING DOOR $E_i$.

| Case | #$E_1$ $\beta_{NE}$ | #$E_1$ $\beta_{BPE}$ | #$E_1$ |
|---|---|---|---|
| E | 153.56 | 175.78 | 329.34 |
| F | 250.78 | 126.16 | 376.94 |
| G | 280.52 | 104.92 | 385.44 |
| Case | #$E_2$ $\beta_{NE}$ | #$E_2$ $\beta_{BPE}$ | #$E_2$ |
| E | 139.44 | 156.22 | 295.66 |
| F | 42,22 | 205,84 | 248,06 |
| G | 12.48 | 227.08 | 236.56 |

TABLE III: 625 INDIVIDUALS DISTRIBUTED EVENLY IN THE BUILDING. 2 EXISTS. EVACUATION TIMES AND MEAN TRAVELLED DISTANCE PER INDIVIDUAL.

| Case | TET (sec) | METxI (m) | MDxI (m) |
|---|---|---|---|
| E | 96.73-106.02 | 14.12-14.13 | 16.94-16.95 |
| F | 121.72-125.75 | 20.35-20.40 | 24.41-24.48 |
| G | 132.40-141.63 | 25.77-25.89 | 30.91-31.08 |

## V. Conclusiones

In this paper we have presented a hybrid simulation model to test behavior designs in an emergency evacuation. The model consists of two sub-models, called environmental (EsM) and pedestrian (PsM). The paper considers that any decision is influenced by the knowledge, perceptions and attitude of people. These aspects lead, even in emergencies, to make and solve a deliberative process. The model pro-poses that the making-decision process that an agent makes to evacuate, will be driven by an engine of behavior which concentrates different mindsets or reactions to face up an emergency situation.

This model along with the computational methodology allows us to build an artificial environment populated with autonomous agents, which are capable of interacting with each other.

The paper develops a series of experiments. First we set up a test environment for an evacuation of individuals whose behavior is out for the predicted nearest door and then for the predicted faster door, both with and without spread of fire and heat. In the second set of experiments, we analyze the evacuation of pedestrian with mixed agent behaviors.

The experiments not only check the impact of individual behavior in the time of evacuation, but also analyze the impact of the parameters that are included in the cost function evaluated by the decision making-process. Among others, we examined the impact of distance, population density and width of the exits.

An important result to notice is the impact of the emergence of a danger zone. When an agent perceives some dynamic factor, such as fire or smoke, it can enter into a process of deliberation of objectives ($E_1$ or $E_2$ in the current experiments) or re-plan the route in case of not being reasonable the chosen plan. The deliberation process is managed by an engine of behaviors that represents different reactions to danger. Note that in all cases, the 625 individuals knew and saw the exit, which optimized its decision making process. In case of ignoring this information, agents can make sub-optimal decisions, increasing not only their MET but the total evacuation time (TET).

## VI. Future Works

One of the main advantages of the proposed model with respect to current CA models is that we can describe a set of individuals with heterogeneous behaviors and provide a framework that will allow us to enhance and add new behaviors to existing ones.

Besides adding new behaviors, our next step is the study and consideration of those events that can cause behavioral changes in the engine. In the model, each agent has a set of psychological (level of stress and memory), physiological (age, sex, speed and level of health) and social (level of training and knowledge) characteristics that describe it, which can also be classified as static (age, sex, etc.) or dynamic (level of health, speed, etc.). It will be necessary to analyze which events on these characteristics, or news arising from the environment sub-model, may cause the changes.

At last, due to the incorporation of new features, the model complexity increases. It knows that a simulation model with computation time and memory requirement scaling linearly with the number of persons or the size of the environment is desirable. In that direction, in the future we will extend the model for supporting high performance simulation in order to accelerate the simulation and take advantage of modern computer architectures.

## ACKNOWLEDGEMENTS

We appreciate the guidance received from the IA research group of National University of San Luis. This research has been supported by the UNSL, Argentina, CONICET Argentina, the MICINN Spain and the MINECO (MICINN) Spain.